# Series solutions of confluent Heun equations in terms of incomplete Gamma-functions

A.M. Ishkhanyan

Institute for Physical Research, NAS of Armenia, 0203 Ashtarak, Armenia

**Abstract.** We present a simple systematic algorithm for construction of expansions of the solutions of ordinary differential equations with rational coefficients in terms of mathematical functions having indefinite integral representation. The approach employs an auxiliary equation involving only the derivatives of a solution of the equation under consideration. Using power-series expansions of the solutions of this auxiliary equation, we construct several expansions of the four confluent Heun equations' solutions in terms of the incomplete Gamma-functions. In the cases of single- and double-confluent Heun equations the coefficients of the expansions obey four-term recurrence relations, while for the bi- and tri-confluent Heun equations the recurrence relations in general involve five terms. Other expansions for which the expansion coefficients obey recurrence relations involving more terms are also possible. The particular cases when these relations reduce to ones involving less number of terms are identified. The conditions for deriving closed-form finite-sum solutions via right-hand side termination of the constructed series are discussed.

**PACS numbers:** 02.30.Gp Special functions, 02.30.Hq Ordinary differential equations, 02.30.Mv Approximations and expansions

**MSC numbers**: 33E30 Other functions coming from differential, difference and integral equations, 34B30 Special equations (Mathieu, Hill, Bessel, etc.), 30Bxx Series expansions

**Keywords:** Confluent Heun equations, special functions, series expansions, multi-term recurrence relations

**Introduction**

The expansions of the solutions of complicated differential equations in terms of simpler mathematical functions have played a notable role in developing the theory of these equations, as well as in their numerous applications in mathematics, physics, engineering, chemistry, biology, etc. Among many well-appreciated examples are the general Heun equation [1] and its four confluent cases [2-4], which are currently widely applied in numerous branches of contemporary physics ranging from atomic, molecular and optical physics to condensed state physics, nuclear physics, astrophysics, cosmology and general relativity [2-4]. The solutions of the Heun equations generalize many known mathematical functions including the hypergeometric, Bessel, Mathieu, Lamé functions, etc., and for this reason the special functions emerging from these equations are supposed to constitute a part of the next generation of mathematical functions [2-4].

A useful step in developing the theory of the Heun equations was the development of expansions of the solutions of these equations in terms of functions more advanced than mere powers. Such expansions have been initiated by Svartholm [5], Erdélyi [6] and continued by Schmidt [7], who used the Gauss hypergeometric functions to construct solutions of the general Heun equation having a wider convergence region as compared with those suggested by the power-series solutions. This useful technique has later been developed to cover many other equations, including the confluent equations of the Heun class [8-25]. Besides, not only the Gauss hypergeometric functions, but a number of other expansion functions have been applied, e.g., the Kummer and Tricomi confluent hypergeometric functions [11-14], Coulomb wave functions [15-17], Bessel and Hankel functions [18], incomplete Beta functions [19-22], Goursat and Appell generalized hypergeometric functions [23-25], and other known special functions.

In the present paper, we consider a simple systematic algorithm for construction of expansions of the solutions $u(z)$ of an ordinary differential equation $L[u(z)] = 0$ in terms of certain functions $\varphi_n(z)$, which have indefinite integral representation. The basic idea is the following. We consider an auxiliary equation involving only the derivatives $d^k u/dz^k$ of a solution $u(z)$ of the equation $L[u] = 0$, but not the function $u(z)$ itself. Next, we introduce a new dependent variable $v(z)$ – weighted first derivative of $u(z)$ according to

$$v(z) = \frac{1}{\Phi(z)} \frac{du}{dz} \qquad (1)$$

with an auxiliary function $\Phi(z)$. Considering now an expansion of $v(z)$ in terms of certain expansion functions $w_n$:

$$v(z) = \sum_{n=0}^{\infty} c_n w_n(z), \qquad (2)$$

and term-by-term integrating Eq. (1) multiplied by $\Phi(z)$, we arrive at an expansion

$$u(z) = C_0 + \sum_{n=0}^{\infty} c_n \varphi_n(z), \quad C_0 = \text{const}, \qquad (3)$$

where
$$\varphi_n(z) = \int \Phi(z) w_n(z) dz. \qquad (4)$$

There are many choices for the weight function $\Phi(z)$ along with the expansion functions $w_n(z)$, for which these integrals are expressed in terms of known special functions. For instance, this takes place if $\Phi(z) = z^\lambda$ and all $w_n(z)$ belong to the class of (generalized) hypergeometric functions. In this way, if $w_n(z)$ is a Kummer confluent hypergeometric function, one arrives at expansions in terms of Goursat generalized hypergeometric functions $_2F_2$ [23], and if $w_n(z)$ is a Gauss hypergeometric function, the result is an expansion in terms of the Clausen generalized hypergeometric functions $_3F_2$ [23].

Probably, the simplest possibility is suggested by the Frobenius-type power-series expansion of $v(z)$ in the neighborhood of point $z_1$ of the complex $z$-plane:

$$v(z) = \sum_{n=0}^{\infty} c_n (z - z_1)^{\mu+n} . \qquad (5)$$

Then, if $\Phi(z)$ is chosen as $e^{\lambda z}$ or $z^\lambda$, the functions $\varphi_n(z)$ turn into the incomplete Gamma- or incomplete Beta-functions, respectively. These functions are defined as

$$\Gamma(a;z) = \int_z^\infty e^{-t} t^{a-1} dt, \quad B(a,b;z) = \int_0^z t^{a-1}(1-t)^{b-1} dt, \qquad (6)$$

with appropriate restrictions imposed on the involved parameters $a,b$, as well as on the allowed variation region of the variable $z$ (see, e.g., [26,27]). Examples of application of this approach for construction of expansions of the solutions of the general and bi-confluent Heun equations in terms of the incomplete Beta-functions are presented in [25, 22].

In the present paper, we apply this approach to construct solutions of the four confluent Heun equations in terms of the incomplete Gamma-functions. Before proceeding to particular developments, it is appropriate to make some notes of general character.

First, we note that the integration constant $C_0$ appearing in the expansion (3) is not arbitrary; rather, it should be appropriately chosen in order to achieve a consistent solution (see the details below). Furthermore, below we assume that the notation $u(z)$ does not refer to a solution normalized to unity at the origin; rather it refers to a solution which is defined with accuracy up to an arbitrary constant factor.

Finally, a useful observation which turns out to be rather helpful in practice, when constructing the expansions, is that when passing to an equation for the new dependent variable $v(z)$ (i.e., weighted first derivative of $u(z)$ according to Eq. (1)) one usually encounters an equation which possesses additional singularities as compared with the starting equation for the function $u(z)$. Since these extra singularities turn out to be regular, one can employ the Frobenius solutions of $v$-equations in the neighborhood of these singularities. In turn, such expansions, after integration as described above, lead to new expansion functions for the series Eq. (3), not used before. For instance, in this way one arrives at expansions in terms of certain generalized hypergeometric functions [28], e.g., in terms of the Appell hypergeometric function $F_1$ of two variables of the first kind [29].

The appearance of an extra singularity in equations obeyed by the functions involving the derivatives of the solutions of the Heun equations have been noticed in several cases [23-25,30-33], in particular, in the context of isomonodromic families of Fuchsian equations parametrized by Painlevé VI solutions. Consider the appearance of these extra regular singular points in more detail.

## 2. Extra singularity of the equation for the derivative of a solution of a Heun equation

Let we have a second-order linear ordinary differential equation with polynomial or rational coefficients $f(z)$ and $g(z)$:

$$\frac{d^2 u}{d z^2} + f(z)\frac{du}{dz} + g(z)u = 0. \tag{7}$$

Singularities of this equation, if any, are involved in functions $f$ and $g$. As it was already mentioned above, here we discuss several relevant equations involving only the derivatives of $u(z)$ (not the function $u(z)$ itself). The first of such equations, a direct one for the first derivative $v = du/dz$, is readily derived by dividing Eq. (7) by $g(z)$, then differentiating the equation and further multiplying it by $g(z)$. The resultant equation reads

$$v_{zz} + \left(f - \frac{g_z}{g}\right)v_z + \left(f_z - f\frac{g_z}{g} + g\right)v = 0, \tag{8}$$

where (and hereafter) the subscripts denote differentiation of corresponding order with respect to the indicated variable. It is seen that because of the term $g_z/g$ new singularities may emerge in this equation as compared with Eq. (7). It is further understood that the additional singularities originate from the zeros of the function $g(z)$, and that the behavior of this function at infinity now interferes with that of the function $f(z)$. Furthermore, since it is the logarithmic derivative $g_z/g = d(\ln g)/dz$ that immediately stands for the appearance of new singularities at finite points of the complex $z$-plane, it is understood that for a polynomial or rational $g(z)$ the new singularities all are regular, regardless of the multiplicity of the roots of the equation $g(z) = 0$. It is also worth to note that further application of the same procedure, this time to Eq. (8), may or may not reveal further new singularities depending on the properties of the coefficients of Eq. (8).

It is interesting that, as long as the Fuchsian differential equations having at the utmost *three* regular singular points are discussed, the described procedure does not lead to new singularities. The situation, however, is changed if Fuchsian equations having more regular singularities are considered. The first such equation is the general Heun equation having *four* regular singular points. In its canonical form, this equation is written as [1]

$$u_{zz} + \left( \frac{\gamma}{z} + \frac{\delta}{z-1} + \frac{\varepsilon}{z-a} \right) u_z + \frac{\alpha\beta z - q}{z(z-1)(z-a)} u = 0. \tag{9}$$

The singular points of this equation, all regular, are $z = 0, 1, a$ and $z = \infty$. For the derivative $v(z) = u_z$ of a solution of this equation, Eq. (8) reads

$$v_{zz} + \left( \frac{\gamma+1}{z} + \frac{\delta+1}{z-1} + \frac{\varepsilon+1}{z-a} - \frac{\alpha\beta}{\alpha\beta z - q} \right) v_z + \frac{\Pi(z)}{z(z-1)(z-a)(\alpha\beta z - q)} v = 0, \tag{10}$$

where $\Pi(z)$ is a quadratic polynomial:

$$\Pi(z) = (1+\alpha)(1+\beta) z (\alpha\beta z - 2q) + (q^2 + q(\gamma + a\gamma + a\delta + \varepsilon) - a\alpha\beta\gamma). \tag{11}$$

As we see, Eq. (10) is a Fuchsian differential equation having in general *five* singular points. Compared with Eq. (9), it has an extra regular singularity located at the (finite or infinite) point $z = q/(\alpha\beta)$. It is indeed understood that this singularity originates from the numerator of the coefficient of the last term in Eq. (9), that is from the root of the above function $g(z)$ which in the case of the general Heun equation is a rational function with a linear in z numerator, namely, $\alpha\beta z - q$. Only in four particular cases, namely, when $q = 0$, $q = \alpha\beta$, $q = a\alpha\beta$ and $\alpha\beta = 0$, this root, $z = q/(\alpha\beta)$, coincides with one of the already existing singular points of Eq. (9), and it is for this reason that only in these four particular cases Eq. (10) presents another general Heun equation with altered parameters [24].

Four confluent modifications of the Heun equation arise via coalescence of some of the singular points of the general Heun equation [2-4]. These equations can be written as

$$P(z) u_{zz} + (\gamma + \delta z + \varepsilon z^2) u_z + (\alpha z - q) u = 0, \tag{12}$$

where $P(z)$ is at most a second-degree polynomial: $P(z) = p_0 + p_1 z + p_2 z^2$ [note that the general Heun equation itself is also written in this form, however, with a cubic polynomial $P(z) = z(z-1)(z-a)$]. If this polynomial has two distinct roots, by shifting the origin and scaling, $z \to s_0 + s_1 z$, it can be written, without loss of generality, as $P(z) = z(z-1)$, and, as a result, we will have the single-confluent Heun equation (SCHE). If the polynomial has a twice multiple root, the latter can be put at the origin: $P(z) = z^2$; we then have the double-confluent Heun equation (DCHE). If $P(z)$ is a linear function, again, by putting its root at the origin, i.e., putting $P(z) = z$, we arrive at the bi-confluent Heun equation (BCHE). Finally, if $P(z)$ is a constant, we have the tri-confluent Heun equation (TCHE).

As it is already mentioned above, the four confluent Heun equations also lead to an extra regular singular point if an equation for the derivative is considered. It is immediately seen from Eq. (8) that in all cases this extra singularity is located at the point $z_0 = q/\alpha$, so that this singularity originates from the accessory parameter $q$ of the predecessor general Heun equation (9). Thus, we conclude that the extra singularity always originates from the last term of the Heun equations. It is shown that the characteristic exponents of this singularity are 0 and 2.

## 3. Heuristic observations leading to incomplete Gamma-function expansions

A common feature of the four confluent Heun equations is that for all of them the infinity is an irregular singular point. The infinity becomes irregularity already in the (single) confluent Heun equation, where the regular singularity of the general Heun equation (9) at $z = a$ has been merged into the regular singularity at infinity. A result of the irregular nature of the singularity of $z = \infty$ is seen in Eq. (12) if it is divided by $P(z)$: all the four equations have a constant term in the coefficient of the first-derivative term. In notations of Eq. (12), this constant is given as $\lambda = \varepsilon, \varepsilon, \delta, \gamma$ for SCHE, DCHE, BCHE, and TCHE, respectively.

Thus, the irregularity of the infinity is pronounced in the second term of the confluent Heun equations, while the extra singularity of the equation for the derivative is exceptionally due to the last term of the equations. Then, to get an insight into how the irregularity of the infinity acts, it is useful to examine the solutions of the truncated confluent Heun equations without the last term: $P(z)u_{zz} + (\gamma + \delta z + \varepsilon z^2)u_z = 0$. The result is written as

$$u(z) = C_1 + C_2 \int e^{-\lambda z} F(z) dz, \quad C_{1,2} = \text{const}, \qquad (13)$$

where $\lambda = \varepsilon, \varepsilon, \delta, \gamma$ and

$$F(z) = (1-z)^{-\delta} z^{-\gamma}, \quad e^{\gamma/z} z^{-\delta}, \quad e^{-\varepsilon z^2/2} z^{-\gamma}, \quad e^{-\delta z^2/2 - \varepsilon z^3/3} \qquad (14)$$

for SCHE, DCHE, BCHE, TCHE, respectively. It is then understood that if we consider a power-series expansion for the function $F(z)$ of the form of Eq. (5), we are immediately being led to an incomplete Gamma-function expansion for $u(z)$:

$$u = C_1 + C_2 \left( \sum_{n=0}^{\infty} c_n \int e^{-\lambda z}(z-z_1)^{\mu+n} dz \right) = C_1 - C_2 \left( \sum_{n=0}^{\infty} \frac{c_n e^{-\lambda z_1}}{\lambda^{1+n+\mu}} \Gamma(1+n+\mu; \lambda(z-z_1)) \right). \quad (15)$$

One more step is to look what happens if one treats the last term of Eq. (12) as a perturbation. Applying, for example, the method of variation of constants, it is readily seen that we again come to similar expansions in terms of the incomplete Gamma-functions.

A further argument supplementing our speculations concerning the appearance of incomplete Gamma-function expansions of the solutions for all the four confluent Heun equations comes from the observation that for some particular specifications of the involved parameters all these equations have particular solutions written in terms of the Kummer confluent hypergeometric function, which allows a "natural" expansion in terms of the incomplete Gamma-functions as follows [27]:

$$_1F_1(a;b;z) = 1 + \frac{a}{b} \sum_{n=0}^{\infty} \frac{(b-a)_n}{(b+1)_n n!} (\Gamma(1+n;-z) - \Gamma(1+n;0)). \quad (16)$$

Thus, there are several cases indicating the incomplete Gamma-function expansions of the solutions of the four confluent Heun equations. Below we show that such expansions are constructed for any set of the involved parameters. The presented approach consists of two steps. First, we pass to an equation obeyed by the weighted derivative function $v = e^{\lambda z} u_z$ and construct a power-series solution of this equation in the neighborhood of its extra (regular) singular point $z_0 = q/\alpha$ (or, if available, another regular singular point; see examples below). Then, the integration produces an incomplete Gamma-function expansion.

A few remarks are now appropriate. The above approach necessarily employs the fact that for the confluent Heun equations the infinity is an irregular singular point. Because of this, the equation for a weighted derivative function has at least one regular and one irregular singular points. This is sufficient for construction of the incomplete Gamma-function expansions, i.e., the structure of the equations does not matter much. Hence, the approach is more general and can be applied to more general equations having an irregular singularity at infinity and potent to produce an additional regular singularity in the equations obeyed by the derivatives of their solutions. In this sense, the situation is somewhat similar to the approach by Svartholm [5] and Schmidt [7], who proposed a regular method for construction of series solutions in terms of the Gauss hypergeometric functions for equations having only two regular singular points in a certain region of the complex $z$-plane. Their method concentrates on these singularities, irrespective of the equations' structure outside the mentioned region containing the two regular singularities.

A similar approach has later been applied by Kurth and Schmidt, who developed a global representation of the solutions of second-order linear differential equations having an irregular singularity of rank one at infinity by series in terms of confluent hypergeometric functions [34]. To this end, it seems appropriate to make a comparison of the incomplete Gamma-functions used in the present paper with the confluent hypergeometric functions applied in [34]. The difference is well seen if one recalls the following representation of the incomplete Gamma-function through the Kummer confluent hypergeometric function [26]

$$\Gamma(a;z) = \Gamma(a) - \frac{z^a}{a} {}_1F_1(a; a+1; -z), \qquad (17)$$

or the representation through the Tricomi confluent hypergeometric function [26,27]

$$\Gamma(a;z) = e^{-z} U(1-a; 1-a; z). \qquad (18)$$

A concluding remark is that in several cases it is possible to construct incomplete Gamma-function expansions, for which the involved expansion functions differently depend on the summation index (see the examples below); then the difference becomes more pronounced.

## 5. Incomplete Gamma-function expansions of types I and II. Discussion

In Appendices 1-4 we present several incomplete Gamma-function expansions of the solutions of the four confluent Heun equations. Note that the forms of the Heun equations adopted here slightly differ from those used in various papers as well as in [2-4]. However, all these forms are readily obtained from those used here by straightforward simple specifications of the involved parameters.

To construct the incomplete Gamma-function expansions, according to the above-described approach, we first write a second-order differential equation for a weighted first derivative of the form

$$v(z) = e^{\Lambda(z)} \frac{du}{dz}, \qquad (19)$$

where $\Lambda(z)$ is a linear, quadratic or cubic polynomial depending on the particular equation at hand. In a certain neighborhood of a point $z = z_1$ of the complex $z$-plane the solution of this equation is expanded into a power series:

$$v(z) = \sum_{n=0}^{\infty} c_n^{(z_1)} (z - z_1)^{\mu+n}, \qquad (20)$$

where the coefficients $c_n^{(z_1)}$ are supposed to be zeros for negative $n$, and the point $z_1$ is a center, around which the expansion is developed.

Now, substitution of $v(z)$ of Eq. (20) into Eq. (19) and subsequent integration results in an incomplete Gamma-function expansion. The particular form of the dependence of the involved incomplete Gamma-functions on the summation index $n$ depends on $\Lambda(z)$. If this is a linear function of $z$, the parameter $a$ of the resultant incomplete Gamma-functions $\Gamma(a_n; s(z-z_1))$ depends on $n$ as $a_n = a_0 + n$ ($a_0 = \text{const}$). We refer to such expansions as of type I. If $\Lambda(z)$ is a quadratic or cubic polynomial, the dependence becomes $a_n = a_0 + n/2$ or $a_n = a_0 + n/3$, respectively. Such expansions are referred to as of type II.

The recurrence relations for the coefficients of the expansions turn out to be of four- or five-term form. In some particular cases the recurrence relations are reduced to ones involving less number of terms. If, by a specification of the involved parameters, it is possible to reduce the relations to two-term ones, then the expansion coefficients are explicitly calculated in terms of the ordinary Gamma-functions and then the solution of the $v$-equation is constructed in terms of the hypergeometric functions. In these cases the solution of the starting confluent Heun equation is finally written either as a quadrature, or, equivalently, as a linear combination, with rational coefficients, of at least two hypergeometric functions. The latter point is readily understood if the solution of a confluent equation of the Heun class is rewritten, using Eq. (12), as

$$u = -\frac{1}{\alpha z - q}\left(P(z)v_z + (\gamma + \delta z + \varepsilon z^2)v\right) \tag{21}$$

with $P(z) = z(z-1)$, $z^2$, $z$, $1$ for single-, double-, bi-, and tri-confluent cases, respectively.

We conclude by noting that the examples of expansions presented below show that incomplete Gamma-function expansions of the solutions of the confluent Heun equations are constructed for any set of the involved parameters with proviso that the parameter $\varepsilon$ standing for the irregular singularity of equations (12) is not zero.

**Acknowledgments**

This research has been conducted within the scope of the International Associated Laboratory (CNRS-France and SCS-Armenia) IRMAS. The research has received funding from the European Union Seventh Framework Programme (FP7/2007-2013) under grant agreement No. 295025 – IPERA. The work has been supported by the Armenian State Committee of Science (SCS Grant No. 13RB-052).

**Appendix 1: (Single) Confluent Heun equation (SCHE)**

This equation has two regular singularities at $z = 0$ and $z = 1$ and an irregular singularity of rank 1 at $z = \infty$:

$$u_{zz} + \left(\frac{\gamma}{z} + \frac{\delta}{z-1} + \varepsilon\right)u_z + \frac{\alpha z - q}{z(z-1)}u = 0. \tag{1.1}$$

The weighted first derivative $v(z) = e^{\varepsilon z}u_z$ obeys the equation

$$v_{zz} + \left(\frac{\gamma+1}{z} + \frac{\delta+1}{z-1} - \varepsilon - \frac{1}{z-z_0}\right)v_z + \frac{\Pi(z)}{z(z-1)(\alpha z - q)}v = 0, \tag{1.2}$$

where $z_0 = q/\alpha$ and $\Pi(z)$ is a quadratic polynomial:

$$\Pi(z) = q^2 + \alpha(\gamma + z\gamma\varepsilon + z^2(\alpha - (\gamma+\delta)\varepsilon) - q(\gamma + \delta + \gamma\varepsilon + z(2\alpha - (\gamma+\delta)\varepsilon)). \tag{1.3}$$

Note that this equation applies also to the limiting case $\alpha = 0$, even though $\Pi(z)$ then becomes a linear function of $z$. Note also that in the latter case Eq. (1.2) presents another single-confluent Heun equation with altered parameters.

Then, for a non-zero $\varepsilon$ we obtain an expansion of type I:

$$u(z) = C_0 - e^{-\varepsilon z_1} \sum_{n=0}^{\infty} \frac{c_n^{(z_1)}}{\varepsilon^{1+n+\mu}} \Gamma(1+n+\mu; \varepsilon(z-z_1)). \tag{1.4}$$

For $z_1 = 0$ the coefficients of the expansion obey the four-term recurrence relation

$$S_n c_n^{(0)} + R_{n-1} c_{n-1}^{(0)} + Q_{n-2} c_{n-2}^{(0)} + P_{n-3} c_{n-3}^{(0)} = 0, \tag{1.5}$$

where 
$$S_n = q(n+\mu)(n+\gamma+\mu), \tag{1.6}$$

$$R_n = q^2 + \alpha\gamma - q(\gamma + \delta + \gamma\varepsilon) + (\alpha(1-\gamma) - q(1+\gamma+\delta+\varepsilon))(n+\mu) - (q+\alpha)(n+\mu)^2, \tag{1.7}$$

$$Q_n = -2q\alpha + (q+\alpha)\gamma\varepsilon + q\delta\varepsilon + (\alpha(\gamma+\delta) + (q+\alpha)\varepsilon)(n+\mu) + \alpha(n+\mu)^2, \tag{1.8}$$

$$P_n = \alpha(\alpha - \varepsilon(n+\gamma+\delta+\mu)). \tag{1.9}$$

For left-hand side termination of the series at $n = 0$ should be $S_0 = 0$, i.e., $\mu = 0$ or $\mu = -\gamma$. The series will terminate from the right-hand side if three successive coefficients vanish for some $N = 1, 2, \ldots$, i.e., if $c_N^{(0)} \neq 0$ and $c_{N+1}^{(0)} = c_{N+2}^{(0)} = c_{N+3}^{(0)} = 0$. From the equation $c_{N+3}^{(0)} = 0$ we find that the termination is possible if $P_N = 0$. For non-zero $\alpha$ this is the case if

$$\alpha = \varepsilon(N + \gamma + \delta + \mu), \quad \mu = 0, -\gamma \tag{1.10}$$

for some $N = 1, 2, \ldots$. The remaining two equations, $c_{N+1}^{(0)} = 0$ and $c_{N+2}^{(0)} = 0$, are then expected to impose two more restrictions on the parameters of the confluent Heun equation. It turns out, however, that in this particular case the two equations lead to a single condition, so that just one additional restriction is imposed on the involved parameters. Indeed, it is checked that some of the roots of the equation $c_{N+1}^{(0)}(q) = 0$ fulfill the second equation too, hence, ensure the termination of the series.

In two particular cases, namely if $\alpha = 0$ or if $q = 0$, the recurrence relation (1.6) is reduced to a three-term one. It is checked that apart from the trivial case $\alpha = q = 0$, when the last term in Eq. (1.1) vanishes, the recurrence relation (1.6) is not reduced to a two-term one.

For $z_1 = z_0$ the coefficients $c_n^{(z_0)}$ again obey a four-term recurrence relation:

$$S_n c_n^{(z_0)} + R_{n-1} c_{n-1}^{(z_0)} + Q_{n-2} c_{n-2}^{(z_0)} + P_{n-3} c_{n-3}^{(z_0)} = 0, \tag{1.11}$$

where
$$S_n = z_0(z_0 - 1)(n + \mu)(n + \mu - 2), \tag{1.12}$$

$$R_n = \gamma - z_0(\gamma + \delta) + \left(1 - 2z_0 - \gamma + z_0(\gamma + \delta) - (z_0 - 1)z_0 \varepsilon\right)(n + \mu) + (2z_0 - 1)(n + \mu)^2, \tag{1.13}$$

$$Q_n = (\gamma - z_0(\gamma + \delta))\varepsilon + (\gamma + \delta + (1 - 2z_0)\varepsilon)(n + \mu) + (n + \mu)^2, \tag{1.14}$$

$$P_n = \alpha - \varepsilon(n + \mu + \gamma + \delta). \tag{1.15}$$

If $z_0 \neq 0, 1$, for left-hand side termination of the series at $n = 0$ should be $S_0 = 0$. For a consistent power series, this is achieved only if $\mu = 2$ (the exponent $\mu = 0$ leads to a logarithmic solution). The series is terminated from the right-hand side for some $N = 1, 2, \ldots$ if $c_{N+1}^{(z_0)} = c_{N+2}^{(z_0)} = 0$ and $P_N = 0$. For non-zero $\alpha$ the last condition is fulfilled if

$$\alpha = \varepsilon(N + \gamma + \delta + \mu), \quad \mu = 2. \tag{1.16}$$

Besides, for termination, the remaining two equations ($c_{N+1}^{(z_0)} = c_{N+2}^{(z_0)} = 0$) should also be satisfied by an appropriate choice of the parameters of the confluent Heun equation.

If $z_0 = 0$ or $z_0 = 1$, i.e. if the extra singularity of the $v$-equation coincides with one of the singularities of the starting confluent Heun equation (1.1), the four-term recurrence relation (1.11) is simplified to a three-term one. Since then the higher-order coefficient $S_n$ identically vanishes for all $n$, the characteristic exponent $\mu$ should fulfill the condition $R_0 = 0$. This leads to $\mu = 1, -\gamma$ if $q = 0$ and $\mu = 1, -\delta$ if $q = \alpha$.

We conclude this appendix by noting that another three-term reduction of the recurrence relation is achieved in the limit $\alpha \to 0$, i.e. if $z_0 \to \infty$ (technically, in this limit the coefficient $P_n$ identically vanishes if one considers Eq. (1.11) multiplied by $\alpha^2$). This is readily understood if we recall that in this case Eq. (1.2) presents another single-confluent Heun equation with altered parameters. Finally, we note that in all three-term cases further reductions to two-term relations are not possible for non-zero $\varepsilon$.

## Appendix 2: Double-confluent Heun equation (DCHE)

For the double-confluent Heun equation, the regular singularities of the general Heun equation (9) at $z = 1$ and $z = a$ are merged separately into the other two to form irregular singularities at, respectively, $z = 0$ and $z = \infty$, each of rank 1:

$$u_{zz} + \left(\frac{\gamma}{z^2} + \frac{\delta}{z} + \varepsilon\right) u_z + \frac{\alpha z - q}{z^2} u = 0. \tag{2.1}$$

Note that in the case $\gamma = 0$ this equation is readily reduced to the confluent hypergeometric equation by the simple change of the dependent variable $u = z^s w(z)$. Hence, we suppose $\gamma \neq 0$. According to the general theory [2-4], this equation has four irreducible parameters. Using a scaling transformation, $z \to s_0 z$, one may fix one of the parameters $\gamma, \varepsilon, \alpha$ to an arbitrary value. Another known case when the solution of Eq. (2.1) is written in terms of the confluent hypergeometric functions (this time, of the argument $\gamma/z$) is the case $\varepsilon = \alpha = 0$. Also, a trivial case is $\alpha = q = 0$ when the last term in Eq. (2.1) vanishes so that the general solution is readily written in quadratures.

The differential equation for the weighted first derivative $v(z) = e^{\varepsilon z} u_z$ is written as

$$v_{zz} + \left(\frac{\gamma}{z^2} + \frac{\delta + 2}{z} - \varepsilon - \frac{1}{z - z_0}\right) v_{zz} + \frac{\Pi(z)}{z^2 (\alpha z - q)} v = 0, \tag{2.2}$$

where $z_0 = q/\alpha$ and $\Pi(z)$ is the quadratic polynomial

$$\Pi(z) = q^2 + q(\gamma\varepsilon - \delta + z(\delta\varepsilon - 2\alpha)) + \alpha\left(z^2(\alpha - \delta\varepsilon) - z\gamma\varepsilon - \gamma\right). \tag{2.3}$$

Accordingly, for a non-zero $\varepsilon$ we obtain an expansion of type I:

$$u(z) = C_0 - e^{-\varepsilon z_1} \sum_{n=0}^{\infty} \frac{c_n^{(z_1)}}{\varepsilon^{1+n+\mu}} \Gamma(1 + n + \mu; \varepsilon(z - z_1)). \tag{2.3}$$

For $z_1 = 0$ the coefficients $c_n^{(0)}$ of this expansion obey the four-term recurrence relation

$$S_n c_n^{(0)} + R_{n-1} c_{n-1}^{(0)} + Q_{n-2} c_{n-2}^{(0)} + P_{n-3} c_{n-3}^{(0)} = 0, \tag{2.4}$$

where
$$S_n = -q\gamma(n + \mu), \tag{2.5}$$

$$R_n = q(q - \delta + \gamma\varepsilon) - a\gamma + (\alpha\gamma - q - q\delta)(n + \mu) - q(n + \mu)^2, \tag{2.6}$$

$$Q_n = \varepsilon(q\delta - \alpha\gamma) - 2q\alpha + (\alpha\delta + q\varepsilon)(n + \mu) + \alpha(n + \mu)^2, \tag{2.7}$$

$$P_n = \alpha\big(\alpha - \varepsilon(n + \mu + \delta)\big). \tag{2.8}$$

If $q\gamma \neq 0$, for left-hand side termination of the series at $n = 0$ should be $S_0 = 0$, hence, the only choice is $\mu = 0$. The series is terminated from the right-hand side for some $N = 1, 2, \ldots$ if $c_{N+1}^{(0)} = c_{N+2}^{(0)} = 0$ and $P_N = 0$. For non-zero $\alpha$, the last condition is fulfilled if

$$\alpha = \varepsilon(n + \mu + \delta), \ \mu = 0. \tag{2.9}$$

Interestingly, it turns out that in this case also the remaining two equations lead to a single condition, namely, some of the roots of the equation $c_{N+1}^{(0)}(q) = 0$ fulfill the equation $c_{N+2}^{(0)}(q) = 0$ too, hence, ensure the termination of the series.

Apart from the degenerate case $\gamma = 0$, the recurrence relation (2.4) is turned into a three-term one if $\alpha = 0$ or $q = 0$. Again, as in the case of SCHE, there are no nontrivial cases, for which the recurrence relation is further reduced to a two-term one.

For $z_1 = z_0$ the coefficients $c_n^{(z_0)}$ also obey a four-term recurrence relation:

$$S_n c_n^{(z_0)} + R_{n-1} c_{n-1}^{(z_0)} + Q_{n-2} c_{n-2}^{(z_0)} + P_{n-3} c_{n-3}^{(z_0)} = 0, \tag{2.10}$$

where
$$S_n = z_0^2 (n + \mu)(n + \mu - 2), \tag{2.11}$$

$$R_n = -(\gamma + z_0 \delta) + \left(\gamma + z_0(\delta - 2) - z_0^2 \varepsilon\right)(n + \mu) + 2z_0(n + \mu)^2, \tag{2.12}$$

$$Q_n = -(\gamma + z_0 \delta)\varepsilon + (\delta - 2z_0 \varepsilon)(n + \mu) + (n + \mu)^2, \tag{2.13}$$

$$P_n = \alpha - \varepsilon(n + \mu + \delta). \tag{2.14}$$

If $z_0 \neq 0$, i.e. if $q \neq 0$, the characteristic exponent $\mu$ should satisfy the equation $S_0 = 0$; hence, $\mu = 0$ or $\mu = 2$. A consistent power series is achieved only for the greater exponent $\mu = 2$. For non-zero $\alpha$, the series terminates from the right-hand side if

$$\alpha = \varepsilon(n + \mu + \delta), \ \mu = 2 \tag{2.15}$$

and $c_{N+1}^{(z_0)} = c_{N+2}^{(z_0)} = 0$, which impose two more restrictions on the parameters of the double-confluent Heun equation (2.1).

If $z_0 = 0$, i.e. if the extra singularity of the $v$-equation coincides with the (irregular) singularity of the starting double-confluent Heun equation (2.1), the four-term recurrence relation (2.10) is simplified to a three-term one. Since then the higher-order coefficient $S_n$ identically vanishes, the characteristic exponent $\mu$ should fulfill the condition $R_0 = 0$. This leads to the only possible exponent $\mu = 1$ for this case.

As in the case of the single-confluent Heun equation, another reduction of the recurrence relation (2.10) to a three-term one is achieved in the limit $\alpha \to 0$ ($z_0 \to \infty$). Finally, we note that in all above three-term cases further reductions to two-term relations are not possible for non-zero $\varepsilon$.

**Appendix 3: Bi-confluent Heun equation (BCHE)**

This equation has a regular singularity at $z = 0$ and an irregular singularity of rank $2$ at $z = \infty$:

$$u_{zz} + \left(\frac{\gamma}{z} + \delta + \varepsilon z\right) u_z + \frac{\alpha z - q}{z} u = 0. \tag{3.1}$$

As it is immediately seen, this equation turns into the Kummer confluent hypergeometric equation if $\varepsilon = 0$ and $\alpha = 0$. In fact, if $\varepsilon = 0$ this equation is always reduced to the confluent hypergeometric equation by the simple change of the dependent variable $u = e^{\lambda z} w(sz)$. Another known case when the solution is written in terms of the confluent hypergeometric functions (this time, of the argument $-\varepsilon z^2/2$) is the case $\delta = q = 0$ (see below Eq. (3.31)). Finally, in a sense trivial is the case $\alpha = q = 0$ when the last term in Eq. (3.1) vanishes so that the general solution is readily written in quadratures.

The solutions of the biconfluent Heun equation allow incomplete Gamma-function expansions of both type I and type II. The expansions of the first type are constructed if one considers the differential equation for the weighted first derivative $v(z) = e^{\lambda(z-z_1)} u_z$:

$$v_{zz} + \left(\frac{\gamma+1}{z} + \delta - 2\lambda + \varepsilon z - \frac{1}{z-z_0}\right) v_z + \frac{\Pi(z)}{z(\alpha z - q)} v = 0, \tag{3.2}$$

where $z_0 = q/\alpha$ and $\Pi(z)$ is the cubic polynomial

$$\Pi(z) = q^2 + q(\gamma\lambda + \lambda - \delta) - \alpha\gamma - \\ (\alpha\gamma\lambda + q(2\alpha + 2\varepsilon - \delta\lambda + \lambda^2))z + (q\varepsilon\lambda + \alpha(\alpha + \varepsilon - \delta\lambda + \lambda^2))z^2 - \alpha\varepsilon\lambda z^3. \tag{3.3}$$

Note that, unlike the above single-confluent and double-confluent cases, here $\lambda$ is an arbitrary non-zero constant, which can be specified, afterwards, as desired.

Accordingly, for a non-zero $\lambda$ we obtain an expansion of type I:

$$u(z) = C_0 - e^{-\lambda z_1} \sum_{n=0}^{\infty} \frac{c_n^{(z_1)}}{\lambda^{1+n+\mu}} \Gamma(1+n+\mu; \lambda(z-z_1)). \tag{3.4}$$

For both $z_1 = 0$ and $z_1 = z_0$ the coefficients $c_n^{(z_1)}$ of this expansion obey a five-term recurrence relation:

$$T_n c_n^{(z_1)} + S_{n-1} c_{n-1}^{(z_1)} + R_{n-2} c_{n-2}^{(z_1)} + Q_{n-3} c_{n-3}^{(z_1)} + P_{n-4} c_{n-4}^{(z_1)} = 0. \tag{3.5}$$

If $z_1 = 0$, the coefficients of this relation are written as

$$T_n = -q(n+\mu)(n+\gamma+\mu), \tag{3.6}$$

$$S_n = q^2 - \alpha\gamma - q(\delta - \lambda - \gamma\lambda) + (\alpha\gamma - \alpha + 2q\lambda - q\delta)(n+\mu) + \alpha(n+\mu)^2, \quad (3.7)$$

$$R_n = -2q(\alpha + \varepsilon) + \lambda(q\delta - \alpha\gamma - q\lambda) + (\alpha\delta - q\varepsilon - 2\alpha\lambda)(n+\mu), \quad (3.8)$$

$$Q_n = \alpha^2 + q\varepsilon\lambda + \alpha\lambda(\lambda - \delta) + \alpha\varepsilon(1+n+\mu), \quad (3.9)$$

$$P_n = -\alpha\lambda\varepsilon, \quad (3.10)$$

where generally $\mu = 0, -\gamma$. Since $P_n$ does not depend on $n$, for nonzero $\alpha\lambda\varepsilon$ the series cannot terminate from the right-hand side.

The recurrence relation (3.5) is reduced to one involving four successive terms if $q = 0$ or $\alpha\varepsilon = 0$ ($\lambda \neq 0$). It is seen from Eq. (3.8) that another four-term reduction, however, involving non-successive terms, is achieved if one chooses $q$ and $\lambda$ so that $R_n \equiv 0$ for all $n$. Besides, if $q = \gamma = 0$, by choosing $\lambda = \delta/2$ the relation is reduced to one involving three (non-successive terms). Similarly, in the case $\varepsilon = 0$ it is possible to specify the constant $\lambda$ so that the relation is reduced to a three-term one: $\lambda^2 - \delta\lambda + \alpha = 0$. This is of course an expected result since, as it was mentioned above, at $\varepsilon = 0$ the biconfluent Heun equation is reduced to the confluent hypergeometric equation. Performing the calculations, one then recovers the above-mentioned expansion of the Kummer hypergeometric function in terms of the incomplete Gamma-functions, Eq. (16).

If $z_1 = z_0$, for the coefficients of the recurrence relation (3.5) we have

$$T_n = z_0(n+\mu)(n+\mu-2), \quad (3.11)$$

$$S_n = \lambda z_0 - (\gamma + \delta z_0 + \varepsilon z_0^2) + (\gamma - 1 + z_0(\delta + z_0\varepsilon - 2\lambda))(n+\mu) + (n+\mu)^2, \quad (3.12)$$

$$R_n = \lambda^2 z_0 - \lambda(\gamma + \delta z_0 + \varepsilon z_0^2) + (\delta + 2z_0\varepsilon - 2\lambda)(n+\mu), \quad (3.13)$$

$$Q_n = \alpha + \varepsilon - \lambda(\delta + 2z_0\varepsilon - \lambda) + \varepsilon(n+\mu), \quad (3.14)$$

$$P_n = -\lambda\varepsilon, \quad (3.15)$$

where generally $\mu = 0, 2$. Here the situation is much similar to the previous case. In this case also the recurrence relation (3.5) is reduced to one involving four successive terms if $q = 0$ or $\varepsilon = 0$ ($\lambda \neq 0$). Another four-term reduction of the recurrence relation (3.5), however, involving non-successive terms, is achieved if $z_0 = -2\gamma/\delta$ and one puts $\lambda = z_0\varepsilon + \delta/2$ ($R_n = 0$). Again, in the case $\varepsilon = 0$ the recurrence relation is reduced to a three-term one by choosing $\lambda$ so that $\lambda^2 - \delta\lambda + \alpha = 0$. Finally, we note that since $P_n$ does not depend on $n$, for nonzero $\varepsilon = 0$ the series cannot terminate from the right-hand side.

Now we present the second type of incomplete Gamma-function expansions of the solutions of the bi-confluent Heun equation (3.1). Several such expansions can be suggested applying the differential equation obeyed by the weighted first derivative of the form $v(z) = z^{\sigma} e^{\lambda z + \tau z^2/2} u_z$. Below we present two examples. The first example is constructed if $v = e^{-\varepsilon z^2/2} u_z$. The equation for $v(z)$ reads

$$v_{zz} + \left(\frac{\gamma+1}{z} + \delta - \varepsilon z - \frac{\alpha}{\alpha z - q}\right) v_z + \frac{\Pi(z)}{z(\alpha z - q)} v = 0, \qquad (3.16)$$

where $\Pi(z)$ is the cubic polynomial

$$\Pi(z) = q^2 - q\delta - \alpha\gamma + (q\gamma\varepsilon - 2q\alpha)z + (q\delta\varepsilon + \alpha(\alpha - \gamma\varepsilon))z^2 - \alpha\delta\varepsilon z^3. \qquad (3.17)$$

Using a Frobenius power-series solution of this equation in the neighborhood of its regular singularity $z = 0$:

$$v(z) = \sum_{n=0}^{\infty} c_n^{(0)} z^{\mu+n}, \qquad (3.18)$$

we get the expansion

$$u(z) = C_0 - \sum_{n=0}^{\infty} \frac{c_n^{(0)}}{2(\varepsilon/2)^{(1+n+\mu)/2}} \Gamma\left(\frac{1+n+\mu}{2}; \frac{\varepsilon z^2}{2}\right), \qquad (3.19)$$

which applies if $\varepsilon \neq 0$. The coefficients $c_n^{(0)}$ obey the five-term recurrence relation

$$T_n c_n^{(0)} + S_{n-1} c_{n-1}^{(0)} + R_{n-2} c_{n-2}^{(0)} + Q_{n-3} c_{n-3}^{(0)} + P_{n-4} c_{n-4}^{(0)} = 0, \qquad (3.20)$$

where
$$T_n = -q(n+\mu)(n+\mu+\gamma), \qquad (3.21)$$

$$S_n = q^2 - \alpha\gamma - q\delta + (\alpha\gamma - \alpha - q\delta)(n+\mu) + \alpha(n+\mu)^2, \qquad (3.22)$$

$$R_n = -2q\alpha + q\gamma\varepsilon + (\alpha\delta + q\varepsilon)(n+\mu), \qquad (3.23)$$

$$Q_n = \alpha^2 + q\delta\varepsilon - \alpha\varepsilon(n+\mu+\gamma), \qquad (3.24)$$

$$P_n = -\alpha\delta\varepsilon, \qquad (3.25)$$

where generally $\mu = 0, -\gamma$. The recurrence relation (3.20) is reduced to one involving four successive terms if $q\delta\alpha = 0$ ($\varepsilon \neq 0$). If $q = 0$ and $\delta = 0$ simultaneously, the recurrence relation becomes two-term:

$$S_{n-1} c_{n-1}^{(0)} + Q_{n-3} c_{n-3}^{(0)} = 0, \qquad (3.26)$$

with $\mu = 0, -1-\gamma$ and ($\alpha \neq 0$)

$$S_n = (n+\mu-1)(n+\mu+\gamma), \quad Q_n = \alpha - \varepsilon(n+\mu+\gamma). \qquad (3.27)$$

Then, the coefficients $c_n^{(0)}$ of the expansion (3.18) are explicitly calculated in terms of the Gamma functions. Using the Pochhammer symbol, the result reads

$$c_n = \frac{1+(-1)^{n-1}}{2} a_{(n-1)/2}, \quad a_k = \frac{(\varepsilon/2)_k \left((1+\gamma+\mu-\alpha/\varepsilon)/2\right)_k}{(1+\mu/2)_k (1+(1+\gamma+\mu)/2)_k}. \quad (3.29)$$

Correspondingly, $v(z)$ is expressed in terms of the confluent hypergeometric functions:

$$v(z) = C_1 z \cdot {}_1F_1\left(-\frac{\alpha}{2\varepsilon}+\frac{1+\gamma}{2};\frac{3+\gamma}{2};\frac{\varepsilon z^2}{2}\right) + C_2 z^{-\gamma} {}_1F_1\left(-\frac{\alpha}{2\varepsilon};\frac{1-\gamma}{2};\frac{\varepsilon z^2}{2}\right). \quad (3.30)$$

This leads to the following general solution of the bi-confluent Heun equation for $\delta = q = 0$:

$$u(z) = C_1 \cdot {}_1F_1\left(\frac{\alpha}{2\varepsilon};\frac{1+\gamma}{2};-\frac{\varepsilon z^2}{2}\right) + C_2 z^{1-\gamma} {}_1F_1\left(\frac{\alpha}{2\varepsilon}+\frac{1-\gamma}{2};\frac{3-\gamma}{2};-\frac{\varepsilon z^2}{2}\right). \quad (3.31)$$

Comparing this solution with the expansion (3.19), after some algebra, we recover the above mentioned expansion of the Kummer confluent hypergeometric function in terms of the incomplete Gamma-functions, Eq. (16).

Apart from the above obvious cases $q\alpha\delta = 0$, another four-term reduction of the recurrence relation (3.20), however, involving non-successive terms, is achieved if one chooses $\gamma\varepsilon - 2\alpha = 0$ and $\alpha\delta + q\varepsilon = 0$ so that $R_n \equiv 0$ for all $n$. Finally, we would like to mention the three-term reduction achieved at $\delta = \alpha = 0$.

A second example of type II expansions can be constructed by means of employing the differential equation obeyed by the function $v = e^{-\varepsilon(z-z_0)^2/2} u_z$:

$$v_{zz} + \left(\frac{\gamma+1}{z} + \lambda + \varepsilon z_0 - \varepsilon z - \frac{1}{z-z_0}\right) v_z + \frac{\Pi(z)}{z(z-z_0)} v = 0, \quad (3.32)$$

where $\lambda = \delta + \varepsilon z_0$ and $\Pi(z)$ is the cubic polynomial

$$\Pi(z) = -\gamma - z_0 \lambda + z_0^2(\alpha - \gamma\varepsilon) - z_0(2\alpha - 2\gamma\varepsilon + z_0\varepsilon\lambda)z + (\alpha - \gamma\varepsilon + 2z_0\varepsilon\lambda)z^2 - \varepsilon\lambda z^3. \quad (3.33)$$

Now, using a Frobenius solution of this equation in the neighborhood of its regular singular point $z = z_0$:

$$v(z) = \sum_{n=0}^{\infty} c_n^{(z_0)} (z-z_0)^{\mu+n}, \quad (3.34)$$

we get the expansion

$$u(z) = C_0 - \sum_{n=0}^{\infty} \frac{c_n^{(z_0)}}{2(\varepsilon/2)^{(1+n+\mu)/2}} \Gamma\left(\frac{1+n+\mu}{2};\frac{\varepsilon(z-z_0)^2}{2}\right), \quad (3.35)$$

which applies if $\varepsilon \neq 0$ and $\alpha \neq 0$ (since $z_0$ should be a finite point of the complex plane).

The coefficients $c_n^{(z_0)}$ obey the five-term recurrence relation

$$T_n c_n^{(z_0)} + S_{n-1} c_{n-1}^{(z_0)} + R_{n-2} c_{n-2}^{(z_0)} + Q_{n-3} c_{n-3}^{(z_0)} + P_{n-4} c_{n-4}^{(z_0)} = 0, \qquad (3.36)$$

where
$$T_n = z_0 (n+\mu)(n+\mu-2), \qquad (3.37)$$

$$S_n = (n+\mu-1)(n+\mu+\gamma+\delta z_0 + \varepsilon z_0^2), \qquad (3.38)$$

$$R_n = \delta(n+\mu), \qquad (3.39)$$

$$Q_n = \alpha - \varepsilon(n+\mu+\gamma+\delta z_0 + \varepsilon z_0^2), \qquad (3.40)$$

$$P_n = -\varepsilon(\delta + \varepsilon z_0), \qquad (3.41)$$

where generally $\mu = 0, 2$. Since $P_n$ does not depend on $n$, the series cannot terminate from the right-hand side unless $\delta + \varepsilon z_0 = 0$ ($\varepsilon = 0$ is forbidden).

It is readily seen that the recurrence relation (3.36) reduces to a four-term one if $z_0 \delta(\delta + \varepsilon z_0) = 0$. Since $z_0 = q/\alpha$, it is understood that further reduction to a three-term one is possible only if $\delta$ and $q$ vanish simultaneously. In this case, however, we have the two-term particular case considered above. Finally, we note that if $z_0^2 + \delta^2 \neq 0$ but $\delta + \varepsilon z_0 = 0$ ($P_n = 0$), we have a series (ruled by a recurrence relation for the coefficients involving three or four terms), which may terminate from the right-hand side. For termination, necessarily should hold $Q_N = 0$ for some $N = 1, 2, \ldots$, that is

$$\alpha = \varepsilon(N + \mu + \gamma), \quad \mu = 2. \qquad (3.42)$$

In addition, in the three-term case one should impose one more restriction ($c_{N+1}^{(z_0)} = 0$), and in the four-term case two such restrictions ($c_{N+1}^{(z_0)} = 0$ and $c_{N+2}^{(z_0)} = 0$) should be fulfilled.

**Appendix 4: Tri-confluent Heun equation**

Here, the singularities $z = 0, 1, a$ of the general Heun equation have coalesced into that at infinity to form an irregular singularity of rank 3 at $z = \infty$:

$$u_{zz} + (\gamma + \delta z + \varepsilon z^2) u_z + (\alpha z - q) u = 0. \qquad (4.1)$$

If $\varepsilon = 0$, this equation is always reduced to simpler equations. If $\varepsilon = 0$ and $\delta \neq 0$, it is reduced to the confluent hypergeometric equation, and it is reduced to the Airy equation if $\varepsilon = \delta = 0$, $\alpha \neq 0$. For this reason, below we suppose $\varepsilon \neq 0$. We note also another known case when Eq. (4.1) is reduced to the confluent hypergeometric equation corresponding to the specification $\gamma = \delta = q = 0$, as well as the case $\alpha = q = 0$ when the last term of the equation vanishes so that the general solution is written in terms of quadratures.

According to the general theory, the tri-confluent Heun equation has only three irreducible parameters. Different specifications of two of the five parameters involved in Eq. (4.1) were applied by different authors depending on the specific theoretical context or particular problem at hand (see, e.g., [2-4]). We note that in the case of non-zero $\varepsilon$ one may fix $\varepsilon$ to any (non-zero) value by scaling $z \to s_1 z$. Simultaneously, in doing this, one may achieve any desired (zero or non-zero) value either for $\gamma$ [4] or $\delta$ [2-3] by shifting the origin: $z \to s_1 z + s_0$. Alternatively, instead of being interested in $\gamma$ or $\delta$, in the case of nonzero $\alpha$, along with fixing the value of $\varepsilon$, one may make the parameter $q$ to adopt any zero or nonzero value. Depending on the particular context of interest, these transformations can be applied to achieve maximal simplifications. For the purposes of the present appendix, it is advantageous to have $q = 0$. However, we use the general form (4.1) since then other forms can be readily employed by straightforward specifications of the involved parameters.

Consider the case $\alpha \neq 0$, so that $z_0 = q/\alpha$ is a finite point of the complex $z$-plane. The differential equation for the function $v(z) = e^{\lambda(z-z_0)} u_z$ is written as

$$v_{zz} + \left(\gamma - 2\lambda + \delta z + \varepsilon z^2 - \frac{1}{z-z_0}\right) v_z + \frac{\Pi(z)}{z-z_0} v = 0, \qquad (4.2)$$

where $\Pi(z)$ is the cubic polynomial

$$\Pi(z) = (\lambda - \gamma - \delta z_0 - \varepsilon z_0^2)(1+\lambda\xi) + (\alpha + \varepsilon - \delta\lambda - 2\varepsilon z_0 \lambda)\xi^2 - \varepsilon\lambda\xi^3, \quad \xi = z - z_0. \qquad (4.4)$$

Accordingly, for a non-zero $\lambda$, applying a Frobenius solution of this equation in the neighborhood of its regular singular point $z = z_0$:

$$v(z) = \sum_{n=0}^{\infty} c_n^{(z_0)} (z-z_0)^{\mu+n}, \qquad (4.3)$$

we derive an expansion of type I:

$$u(z) = C_0 - \sum_{n=0}^{\infty} \frac{c_n^{(z_0)}}{\lambda^{1+n+\mu}} \Gamma(1+n+\mu; \lambda(z-z_0)), \qquad (4.4)$$

the coefficients $c_n^{(z_0)}$ of which obey the five-term recurrence relation

$$T_n c_n^{(z_0)} + S_{n-1} c_{n-1}^{(z_0)} + R_{n-2} c_{n-2}^{(z_0)} + Q_{n-3} c_{n-3}^{(z_0)} + P_{n-4} c_{n-4}^{(z_0)} = 0, \qquad (4.5)$$

with

$$T_n = (n+\mu)(n+\mu-2), \qquad (4.6)$$

$$S_n = -\gamma - z_0(\delta + \varepsilon z_0) + \lambda + (\gamma + z_0(\delta + \varepsilon z_0) - 2\lambda)(n+\mu), \qquad (4.7)$$

$$R_n = (-\gamma - z_0(\delta + \varepsilon z_0) + \lambda)\lambda + (\delta + 2\varepsilon z_0)(n + \mu), \tag{4.8}$$

$$Q_n = \alpha + \varepsilon - (\delta + 2\varepsilon z_0)\lambda + \varepsilon(n + \mu), \tag{4.9}$$

$$P_n = -\varepsilon\lambda, \tag{4.10}$$

where one should choose the greater exponent $\mu = 2$.

The only case when the recurrence relation (4.5) involves four terms (however, non-successive) is achieved if $\delta + 2z_0\varepsilon = 0$ and one chooses $\lambda = \gamma - \varepsilon z_0^2$, since then the coefficient $R_n$ vanishes for all $n$. We note that the quantity $\delta + 2z_0\varepsilon$ is not changed by the scaling and shifting transformation $z \to s_1 z + s_0$, hence, the recurrence relation necessarily involves five terms unless $\delta + 2z_0\varepsilon = 0$. Finally, we note that since $P_n$ does not depend on $n$, the series cannot terminate from the right-hand side.

Considering the differential equation for the function $v(z) = e^{\lambda(z-z_0)^2/2} u_z$:

$$v_{zz} + \left(\gamma + 2z_0\lambda + (\delta - 2\lambda)z + \varepsilon z^2 - \frac{1}{z - z_0}\right) v_z + \frac{\Pi(z)}{z - z_0} v = 0, \tag{4.11}$$

where $\Pi(z)$ is the quartic polynomial

$$\Pi(z) = -(\gamma + \delta z_0 + \varepsilon z_0^2)(1 + \lambda\xi^2) + (\alpha + \varepsilon)\xi^2 - \lambda(\delta + 2\varepsilon z_0 - \lambda)\xi^3 - \varepsilon\lambda\xi^4, \quad \xi = z - z_0, \tag{4.12}$$

and applying the Frobenius solution of this equation in the neighborhood of its regular singular point $z = z_0$, we now obtain an expansion of type II:

$$u(z) = C_0 - \sum_{n=0}^{\infty} \frac{c_n^{(z_0)}}{2(\lambda/2)^{(1+n+\mu)/2}} \Gamma\left(\frac{1+n+\mu}{2}; \frac{\lambda(z-z_0)^2}{2}\right), \tag{4.13}$$

the coefficients $c_n^{(z_0)}$ of which, however, obey a six-term recurrence relation:

$$W_n c_n^{(z_0)} + T_{n-1} c_{n-1}^{(z_0)} + S_{n-2} c_{n-2}^{(z_0)} + R_{n-3} c_{n-3}^{(z_0)} + Q_{n-4} c_{n-4}^{(z_0)} + P_{n-5} c_{n-5}^{(z_0)} = 0, \tag{4.14}$$

with

$$W_n = (n + \mu)(n + \mu - 2), \tag{4.15}$$

$$T_n = (\gamma + z_0(\delta + \varepsilon z_0))(n + \mu - 1), \tag{4.16}$$

$$S_n = (\delta + 2\varepsilon z_0 - 2\lambda)(n + \mu), \tag{4.17}$$

$$R_n = \alpha + \varepsilon - (\gamma + z_0(\delta + \varepsilon z_0))\lambda + \varepsilon(n + \mu), \tag{4.18}$$

$$Q_n = -(\delta + 2\varepsilon z_0 - \lambda)\lambda, \tag{4.19}$$

$$P_n = -\varepsilon\lambda, \tag{4.20}$$

where one should put $\mu = 2$. As in the previous case, since $P_n$ does not depend on $n$, this series cannot terminate from the right-hand side.

Inspecting the coefficients $S_n$ and $Q_n$, we see that if $\delta + 2z_0\varepsilon \neq 0$, one can reduce the recurrence relation (4.14) to one involving five terms (however, non-successive) by putting either $2\lambda$ or $\lambda$ equal to $\delta + 2z_0\varepsilon$, thus forcing, respectively, $S_n \to 0$ or $Q_n \to 0$. If in addition $z_0$ is a root of the quadratic equation $\gamma + \delta z_0 + \varepsilon z_0^2 = 0$, the coefficient $T_n$ also vanishes and the recurrence relation involves only four terms. If, however, $\delta + 2z_0\varepsilon = 0$, since $\lambda$ cannot be zero, the recurrence relation (4.14) may be reduced at most to a five-term one (achieved when $\gamma = 0$, i.e. $T_n = 0$).

We conclude by discussing one more expansion constructed when the differential equation for the function $v(z) = e^{\varepsilon(z-z_0)^3/3} u_z$:

$$v_{zz} + \left(\gamma - 2\varepsilon z_0^2 + (\delta + 4\varepsilon z_0)z - \varepsilon z^2 - \frac{1}{z - z_0}\right) v_z + \frac{\Pi(z)}{z - z_0} v = 0, \qquad (4.21)$$

where $\Pi(z)$ is the quartic polynomial

$$\Pi(z) = -(\gamma + \delta z_0 + \varepsilon z_0^2)(1 + \varepsilon\xi^3) + \alpha\xi^2 - \varepsilon(\delta + 2z_0\varepsilon)\xi^4, \quad \xi = z - z_0. \qquad (4.22)$$

Applying the Frobenius solution of this equation in the neighborhood of the regular singular point $z = z_0$, we obtain another expansion of type II:

$$u(z) = C_0 - \sum_{n=0}^{\infty} \frac{c_n^{(z_0)}}{3(\varepsilon/3)^{(1+n+\mu)/3}} \Gamma\left(\frac{1+n+\mu}{3}; \frac{\varepsilon(z-z_0)^3}{3}\right), \qquad (4.23)$$

the coefficients $c_n^{(z_0)}$ of which obey the six-term recurrence relation

$$W_n c_n^{(z_0)} + T_{n-1} c_{n-1}^{(z_0)} + S_{n-2} c_{n-2}^{(z_0)} + R_{n-3} c_{n-3}^{(z_0)} + Q_{n-4} c_{n-4}^{(z_0)} + P_{n-5} c_{n-5}^{(z_0)} = 0, \qquad (4.24)$$

with
$$W_n = (n + \mu)(n + \mu - 2), \qquad (4.25)$$

$$T_n = (\gamma + z_0(\delta + \varepsilon z_0))(n + \mu - 1), \qquad (4.26)$$

$$S_n = (\delta + 2\varepsilon z_0)(n + \mu), \qquad (4.27)$$

$$R_n = \alpha - \varepsilon(n + \mu), \qquad (4.28)$$

$$Q_n = -\varepsilon(\gamma + z_0(\delta + \varepsilon z_0)), \qquad (4.29)$$

$$P_n = -\varepsilon(\delta + 2\varepsilon z_0), \qquad (4.30)$$

where generally $\mu = 0, 2$. As in the previous two cases, since $P_n$ does not depend on $n$, in general this series cannot terminate from the right-hand side.

As it is immediately seen, if $\delta + 2z_0 \varepsilon = 0$ or $\gamma + \delta z_0 + \varepsilon z_0^2 = 0$, the recurrence relation involves four terms. If these two equations are satisfied simultaneously, i.e. if $\gamma = \varepsilon z_0^2 = \varepsilon q^2 / \alpha^2$ and $\delta = -2z_0 \varepsilon = -2\varepsilon q / \alpha$, we have a two-term recurrence relation:

$$W_n c_n^{(z_0)} + R_{n-3} c_{n-3}^{(z_0)} = 0, \tag{4.31}$$

with 
$$W_n = (n+\mu)(n+\mu-2), \quad R_n = \alpha - \varepsilon(n+\mu), \quad \mu = 0, 2. \tag{4.32}$$

The coefficients $c_n^{(z_0)}$ of the expansion (4.23) are then explicitly calculated in terms of the Gamma functions. The result for $\mu = 0$ and initial conditions $c_0^{(z_0)} = 1$, $c_1^{(z_0)} = c_2^{(z_0)} = 0$ reads

$$c_n^{(z_0)} = \frac{1 + 2\cos(2\pi n/3)}{3} \frac{(\varepsilon/3)^{n/3} (-\alpha/(3\varepsilon))_{n/3}}{(1/3)_{n/3} (1)_{n/3}}. \tag{4.33}$$

Correspondingly, the auxiliary function $v(z)$ is expressed in terms of the confluent hypergeometric functions as

$$v(z) = C_1 z^2 {}_1F_1\left(\frac{2}{3} - \frac{\alpha}{3\varepsilon}; \frac{5}{3}; \frac{\varepsilon(z-z_0)^3}{3}\right) + C_2 \cdot {}_1F_1\left(-\frac{\alpha}{3\varepsilon}; \frac{1}{3}; \frac{\varepsilon(z-z_0)^3}{3}\right). \tag{4.34}$$

This leads to the following general solution of the tri-confluent Heun equation for $\gamma = \varepsilon q^2 / \alpha^2$ and $\delta = -2\varepsilon q / \alpha$:

$$u(z) = C_1 \cdot {}_1F_1\left(\frac{\alpha}{3\varepsilon}; \frac{2}{3}; -\frac{\varepsilon(z-z_0)^3}{3}\right) + C_2 z \cdot {}_1F_1\left(\frac{\alpha}{3\varepsilon} + \frac{1}{3}; \frac{4}{3}; -\frac{\varepsilon(z-z_0)^3}{3}\right). \tag{4.35}$$

Of course, this solution is readily deduced from the known one for $\gamma = \delta = q = 0$.

24equation and symmetries of the Painlevé PVI equation", Theor. Math. Phys. **155**, 722 (2008).
31. S.Y. Slavyanov, "Relations between linear equations and Painlevé's equations", Constr. Approx. **39**, 75 (2014).
32. K. Takemura, "Integral representation of solutions to Fuchsian system and Heun's equation" J. Math. Anal. Appl. **342**, 52 (2008).
33. K. Takemura, "Middle Convolution and Heun's Equation", SIGMA **5**, 040 (2009).
34. Th. Kurth and D. Schmidt, "On the global representation of the solutions of second-order linear differential equations having an irregular singularity of rank one in by series in terms of confluent hypergeometric functions", SIAM J. Math. Anal. **17**, 1086 (1986).